\documentclass[12pt,preprint]{emulateapj}
\usepackage{epsfig}
\shorttitle{Spectroscopic abundance analysis of IC\,4665 dwarfs}
\shortauthors{Shen, Jones, Lin, Liu, \& Li}

\begin{document}
\title{Spectroscopic abundance analysis of dwarfs in young open cluster IC\,4665}

\author{Z.-X. Shen}
\affil{Department of Astronomy, Peking University, Beijing 100871,
        P. R. China;
{shenzx@bac.pku.edu.cn}}
\and

\author{B. Jones}
\affil{ UCO/Lick Observatory, Departments of Astronomy and Astrophysics, University
    of California, Santa Cruz CA 95064, U.S.A.}
\and

\author{D. N. C. Lin}
\affil{ UCO/Lick Observatory, Departments of Astronomy and Astrophysics, University
    of California, Santa Cruz CA 95064, U.S.A.}
\and

\author{X.-W. Liu}
\affil{Department of Astronomy, Peking University, Beijing 100871,
        P. R. China; liuxw@bac.pku.edu.cn}
\and

\author{S.-L. Li}
\affil{Department of Astronomy, Peking University, Beijing 100871,
        P. R. China}


\begin{abstract} 
We report a detailed spectroscopic abundance analysis for a sample of 18 F-K
dwarfs of the young open cluster IC\,4665.  Stellar parameters and element
abundances of Li, O, Mg, Si, Ca, Ti, Cr, Fe and Ni have been derived using the
spectroscopic synthesis tool SME (Spectroscopy Made Easy).  Within the
measurement uncertainties the iron abundance is uniform with a standard
deviation of 0.04\,dex. No correlation is found between the iron abundance and
the mass of the stellar convective zone, and between the Li abundance and the
Fe abundance.  In other words, our results do not reveal any signature of
accretion and therefore do not support the scenario that stars with planets
(SWPs) acquire their on the average higher metallicity compared to field stars
via accretion of metal-rich planetary material. Instead the higher metallicity
of SWPs may simply reflect the fact that planet formation is more efficient in
high metallicity environs.  However, since that many details of the planet
system formation processes remain poorly understood, further studies are
needed for a final settlement of the problem of the high metallicity of
SWPs.

The standard deviation of [Fe/H] deduced from our observations, taken as an
upper limit on the metallicity dispersion amongst the IC\,4665 member stars,
has been used to constrain proto-planetary disk evolution, terrestrial and
giant planets formation and evolution processes. The total reservoir of heavy
elements retained by the nascent disks is limited and high retention efficiency
of planet building material is supported. Under modest surface density, gas
giant planets are expected to form in locally enhanced regions or start
efficient gas accretion when they only have a small core of a few $M_\oplus$.
Our results do not support the possibility that the migration of gas giants and
the circularization of terrestrial planets' orbits are regulated by their
interaction with a residual population of planetesimals and dust particles. 

\end{abstract}


\section{Introduction} 

Since the first report of an extrasolar planet orbiting a solar-type star in
1995 (Mayor \& Queloz 1995), over one hundred planetary systems have been
discovered. Two competing scenarios of planet formation and evolution are
gravitational instability and core accretion-gas capture. The theories can be
discriminated and tested by studying the observed properties of extra solar
planetary systems (see e.g.  review by Bodenheimer \& Lin 2002).  A key
property of stars with planets (SWPs) that has emerged from the extensive
observations hitherto has been that the metal abundance of SWPs is on the
average about 0.25\,dex higher than those of typical control sample stars (Laws
et al. 2003) and that the frequency of planets is a strong function of the
metallicity of the host star (Santos et al. 2004; Fischer et al. 2004).  Two
possible explanations have been put forward: either planets form preferentially
around stars rich in heavy elements or the observed overabundance is caused by
enrichment as a result of accretion of H-deficient planetary material onto the
stellar photosphere (Gonzalez 1997).

In contrast to Jupiter and Saturn, a number of extra solar planets are found to
have orbital periods of several days.  Orbital migration predicted by the
current models of planet formation has been invoked to explain the occurrence
of close-in giant planets found by radial velocity surveys (e.g., Lin,
Bodenheimer \& Richardson 1996). As the planet migrates, it scatters along the
way planetesimals in the disk, causing them to fall onto the central star and
enrich the stellar surface layer with heavy elements.  Scattering amongst
planetesimals and angular momentum losing of giant planets, also leads to the
bombardment of their host star.  Material fallen onto the star is diluted in
the convection zone (CZ). Thus depending on the mass of the CZ, a given amount
of accreted planetary material will lead to different level of overabundances
of the stellar surface layer in heavy elements (see e.g. Gonzalez 1997). For a
given spectral type (i.e.  initial mass), the CZ mass decreases from the
initial value of the entire stellar mass in the early stage of contraction to a
fraction as it approaches and passes the zero-age-main-sequence.  For solar
type stars, the decline of the CZ mass ceases after about 30 Million years,
reaching a nominal value of around 0.025\,$M_\odot$. Accretion of
metal-enriched protoplanetary material after this epoch is potentially
detectable.  Note that for solar type main sequence stars, CZ mass decreases as
stellar mass ($M_\ast$) increases. The time scale for reaching the main
sequence also decreases with increasing $M_\ast$.  For a given mass of
protoplanetary material, any signature of the protracted accretion should
therefore be more evident in stars of early spectral types.

Open clusters are physically related groups of stars which are believed to form
from the same homogeneous large cloud in the Galaxy at almost the identical
time.  Therefore one expects that stars of a given young cluster should all
have the same surface chemical abundances, unless they are altered by some
physical processes. These remarkable properties make young clusters excellent
testbeds to test the scenario of accretion of H-deficient planetary material as
the cause of the high metallicity of SWPs. Signature that accretion has
occurred can be revealed by searching for star-to-star variations in metal
content.  In addition, in this scenario, one expects that the enrichment will
be more pronounced in F dwarfs than in stars of later types, since, as
discussed above, F-type dwarfs have a smaller CZ.
 
Several spectroscopic studies aimed at detecting signatures of pollution by
accretion of planetary material by measuring the metallicity dispersion in open
clusters have been published.  Wilden et al. (2002) observed 16 stars in the
Pleiades and found evidence indicating possible accretion in one star, which
shows an excessive metallicity 0.1\,dex above the cluster mean.  Paulson et al.
(2003) analyzed 55 FGK dwarfs in the Hyades and they found two stars of
abundances 0.2\,dex in excess of the cluster mean. Unfortunately, the cluster
membership of the two stars turns out to be questionable.  Similar studies have
also been extended to binary or multiple star systems (Laws \& Gonzalez 2001;
Gratton et al. 2001; Desidera et al.  2004).  Gratton et al. (2001) carry out
a differential abundance analysis for six main sequence binary systems and
found one pair (HD\,219542) exhibiting a 0.09\,dex iron content difference. A
more recent study of this system by Desidera et al. (2004) however shows that
the result of Gratton et al. is probably spurious.  Laws \& Gonzalez (2001)
report the detection of a small metallicity difference between the two
components of 16\,Cyg.  Desidera et al. (2004) perform differential abundance
analysis for 23 wide binaries. They find that most of the pairs exhibit
abundance differences smaller than 0.02\,dex and no pairs show differences in
excess of 0.07\,dex.

Murray et al. (2001) suggest that lithium serves as a good tracer of the mixing
depth of the convective zone. $^6$Li is easily destroyed during the
pre-main-sequence evolutionary phase and should therefore be completely absent
in main sequence stars. If any $^6$Li is detected in mature solar type stars,
there would be a high probability that it is accreted.  The recent claims of
detection of $^6$Li in SWP HD\,82943 by Israelian et al. (2001; 2003), are
however disputed by Reddy et al. (2002; but see also Israelian et al. 2004a).
The absence of $^6$Li in other SWP's however suggests such events are rare
among {\it mature} stars (Mandell et al. 2004).  Nevertheless, if $^6$Li
depletes within a few Myr after the planet consumption, this signature would be
erased among the main sequence stars. Compared to $^6$Li, destruction of the
more abundant isotope $^7$Li requires a higher temperature of 2.5 million
Kelvin. If the temperature at bottom of the convective envelope exceeds this
value, then any $^7$Li in the envelope will be destroyed as well.  For stars
with a thick convective zone (type G or later), $^7$Li is destroyed within a
few hundred million years.  Stars more massive than late F have much thinner
convective envelopes where the base temperatures are generally too low to
destroy lithium.  Consequently $^7$Li can survive for a few billion years in
those stars.  However, in addition to the so-called "lithium-dip" around
6500\,K, generally attributed to some extra mixing mechanism, scatters have
also been observed at all other temperatures in many clusters. The cause of the
lithium abundance dispersion amongst stars of the same temperature has long
been debated. If accretion plays a role in this problem, then one expects that
a dispersion in iron abundance should also be observable (Israelian et al.
2004b).

In this paper, we present spectroscopic abundance analysis for a sample of 18 F
to early K dwarfs in the open cluster IC\,4665, using spectra obtained with the
Keck\,I 10\,m telescope.
IC\,4665 ($l = 30.62$, $b = +17.08$\,deg), at a distance of 350\,pc, is a young
open cluster, but not so young as the T Tauri stars or other very young
associations. Mermilliod (1981a, b) estimates an age of 3--$4\times 10^7$\,yr
and a reddening of $E(B-V)=0.18$.  Allain et al. (1996) confirm that the age of
IC\,4665 is close to $\alpha$ Per ($\sim 50$\,Myr) by studying the rotational
periods and star-spot activities of young solar-type dwarfs in IC\,4665.  The
age is particularly meaningful for the current study aimed at detecting
possible variations of abundance dispersion with the CZ mass: after this age,
the mass of the CZ stays essentially constant for solar mass stars.  The star
membership of IC\,4665 has been studied by Prosser (1993) and Prosser \&
Giampapa (1994), utilizing astrometric, photometric and spectroscopic data.
Being a very young cluster, stellar surface activities in this cluster
(starspots, coronal X-ray emission and H$\alpha$ emission) have been widely
studied (e.g., Allian et al. 1996; Giampapa, Prosser \& Fleming 1998;
Mart\'{i}n \& Montes 1997; Messina et al. 2003). No detailed spectroscopic
abundance analysis has been reported, however, except the study of Li
abundances in 14 dwarfs by Mart\'{i}n \& Montes (1997).   

Since clusters are generally far from us, obtaining high quality spectra for
large samples of cluster members have only become feasible in recent years with
the advent of 10\,m class telescopes.  Except for Fe and Li, only very
restricted measurements of abundances of other elements in cluster stars have
been published. Available data include detailed abundance analysis for 12
Hyades stars (Cayrel et al. 1985), 16 Pleiades stars (Wilden et al.  2002), 9
M\,34 stars (Schuler et al. 2003), 22 IC\,4651 stars (Pasquini et al. 2004) and
a few Pleiades and NGC\,2264 stars (King et al.  2000). Nevertheless, given
that the initial stellar mass is the only variable parameter in a cluster,
detailed cluster elemental abundance analysis can potentially not only place
strong constraints on the enrichment of the interstellar medium, the star
formation history in the disk, but can also provide excellent testbeds for the
stellar evolution theory. It is expected that the current work may prove useful
for this endeavour by adding a substantial amount of new data. 
 
The observations and data reduction procedures are described in Section~2.
Section~3 describes abundance analysis using the spectral synthesis tools SME
(Spectroscopy Made Easy) and presents the results. It is followed by an error
analysis in Section~4. Abundance dispersions yielded by our observations are
discussed in Section~5. The implications of our results on the planet formation
processes are discussed in Section~6 and on protoplanetary disk masses in
Section~7. We conclude with a brief summary in Section~8.

\section{Observation and data reduction}

The spectra were obtained in October 1999 and October 2000 using the HiRes
spectrograph (Vogt 1992) mounted on the Keck\,I 10\,m telescope.  A Tektronix
$2048 \times 2048$ CCD of 24 $\times 24\,\mu{\rm m}$ pixel size was used as the
detector. The spectra covered the wavelength range from 6300 to 8730\,\AA,
split into 16 orders, with small inter-order gaps amongst them.  The
integration time ranged from 10 minutes to half an hour, yielding
signal-to-noise ratios from $\sim 30$ to 150 per resolution element, at a
resolving power of about 60,000. 

The spectra were reduced using IRAF following the procedures described in
Soderblom et al. (1993). The noao.imred.echelle package was used for
flat-fielding, scattered light removal and order extraction. Wavelength
calibration was achieved using exposures of a Th-Ar lamp.

\section{Abundance determinations with SME}

Spectral analyses, including determinations of the stellar parameters and
elemental abundances, were carried using the software package SME (Spectroscopy
Made Easy), originally developed by Valenti \& Piskunov (1996).  SME determines
the basic stellar parameters and elemental abundances by matching the
synthesized spectrum to the observed one. It uses Kurucz stellar model
atmospheres and solves radiative transfer to create synthetic spectra.  A
nonlinear least squares algorithm is then used to solve for any subset of the
input parameters, including $T_{\rm eff}$, $\log\,g$, radial and rotational
velocities, micro- and macro-turbulence velocities and element abundances. The
radiative transfer routine in SME assumes LTE and negligible magnetic field,
and neglects molecular line opacity. Due to these limitation, the current
analysis is limited to stars of spectral types F, G, and early K. Stars of
later types have been excluded.

\begin{table}
\caption{\label{sun} SME solar model parameters}
\centering
\begin{tabular}{lr}
\hline
\hline
\noalign{\smallskip}
Parameter & Value~~~~\\
\noalign{\smallskip}
\hline
\noalign{\smallskip}
$T_{\rm eff}$ & 5735\,K\\
$\log\,g$ & 4.44\,cm\,s$^{-2}$\\
{[M/H]} & 0.0\\
$v_{\rm mic}$ & 0.73\,km\,s$^{-1}$\\
$v_{\rm mac}$ & 3.1\,km\,s$^{-1}$\\
$V\sin{i}$ & 1.60\,km\,s$^{-1}$\\
$\Delta\Gamma_6$ & 2.5\\
\noalign{\smallskip}
\hline
\end{tabular}
\end{table}

Atomic data of spectral lines ($\log\,gf$, van der Waals damping constants)
were initially retrieved from the Vienna Atomic Line Data-base (VALD, Piskunov
et al. 1995; Ryabchikova et al. 1999; Kupka et al. 1999;) using the "stellar" requests with the
parameter expected fractional depth set to greater than 0.2\,\%.  Seven
spectral segments, each of approximately 20 \,\AA\, wide, were selected,
centered at 6336, 6425, 6490, 6595, 6831, 8435 and 8708\,\AA. The segments
include \ion{Fe}{2} and \ion{Fe}{1} lines from a broad range of equivalent
widths, excitation potentials and $\log\,gf$ values to allow accurate stellar
parameters to be derived. Additional segments of widths less than 10\,\AA\,
were then selected for the purpose of abundance determinations for individual
elements, e.g. one centered at 6707\,\AA\, for Li, one at 7774\,\AA\,
for O, and another at 7230\,\AA\, for Si, etc.

In the analysis, we have followed the procedures described in Wilden et al.
(2002).  Firstly a model solar spectrum was created using the atomic data
retrieved from the VALD and compared to the high-resolution, high
signal-to-noise ratio (SNR) solar spectrum (Kurucz et al. 1984) from the National Solar Observatory
(NSO). Bad spectral regions were masked out. We then used the solar spectrum to
solve for improved atomic data of spectral lines of interest in the
pre-selected spectral segments.  The improved atomic data thus obtained were
later used to determine basic stellar parameters as well as elemental
abundances for our sample stars in IC\,4665. 

The determination of solar photospheric abundances depends on stellar
atmospheric models. In an innovative yet controversial approach, Asplund et al.
(2005) have recently published a new solar abundance scale deduced based on a
time-dependent, 3D hydrodynamical model of the solar atmosphere. Given that the
current work is more interested in the relative abundances (or more precisely
the abundance dispersion amongst the cluster members) and the fact that SME
uses a 1D stellar atmosphere, we have opted to use the `old' solar abundance
scale determined based on the traditional 1D model by Grevesse, Noels \& Sauval
(1996). The basic stellar parameters of the Sun derived for the SME solar
model are listed in Table~\ref{sun}. The parameter $\Delta\Gamma_6$ in the list
refers to the enhancement factor of the van der Waals damping constant. These
values were adopted when determining $\log\,gf$ values of spectral lines. To
reduce the number of free parameters, the solar surface gravity was fixed at
the standard value of 4.44\,cm\,s$^{-2}$. The overall metallicity parameter,
[M/H], was used to interpolate the grid of model atmospheres and to scale the
solar elemental abundances (except for elements to be solved) when calculating
opacities.  In addition to microturbulent velocity $v_{\rm mic}$,
macroturbulent velocity $v_{\rm mac}$ and rotation velocity $V\sin{i}$ were
also set individually.  However, since it is difficult to distinguish the
effects of rotation and macroturbulence, especially when fitting line profiles
of slowly rotating stars such as the Sun, usually we fixed one parameter and
solved for the other.  SME ignores stellar surface differential rotation and
treats the star as a uniform rotator.  The simplification does not pose a
problem for slowly rotating stars.  After solving the free atomic parameters
($\log\,gf$ and the van der Waals damping constants), our best fit solar model
spectrum yields a Marquardt algorithm $\chi^2$ value of 180 and a line rms of
0.8\,\%, compared to the corresponding values of 353 and 1.19\,\%,
respectively, obtained by  Wilden et al. (2002). 

 \begin{table*}
 \tabcolsep 10pt
 \centering
 \caption{\label{para} Stellar parameters}
 \begin{tabular}{lcccccc}
 \hline
 \hline
 \noalign{\smallskip}
 Star& $T_{\rm eff}$(SME) & $\log\,g^{\mathrm{a}}$ & $v_{\rm mic}$ & $v_{\rm mac}$$^{\mathrm{b}}$ & $V\sin{i}$ & $V\sin{i}^{\mathrm{c}}$ \\
          & (K) &(cm\,s$^{-2}$) & (km\,s$^{-1}$) & (km\,s$^{-1}$) & (km\,s$^{-1}$)& (km\,s$^{-1}$)\\
 \noalign{\smallskip}
 \hline
 \noalign{\smallskip}
P\,19                 &  6370 & 4.435 & 0.26 & 4.54 & 5.20 & $<$10 \\  
P\,147                &  6189 & 4.490 & 1.02 & 3.97 & 4.21 &     * \\
P\,39$^{\mathrm{d}}$  &  5867 & 4.503 & 1.25 & 3.85 & 13.6 &    15 \\
P\,107                &  5626 & 4.560 & 0.68 & 3.36 & 30.3 &    27 \\
P\,150$^{\mathrm{d}}$ &  5535 & 4.572 & 1.62 & 3.35 & 26.6 &    25 \\
P\,151                &  5494 & 4.583 & 1.46 & 3.15 & 12.4 &     * \\
P\,60                 &  5483 & 4.583 & 1.27 & 3.15 & 9.83 &    13 \\
P\,199                &  5168 & 4.641 & 0.57 & 2.55 & 3.17 &     * \\
P\,75$^{\mathrm{d}}$  &  5347 & 4.465 & 1.60 & 3.10 & 14.1 &    16 \\
P\,165                &  5292 & 4.594 & 1.56 & 3.05 & 31.8 &    40 \\
P\,267                &  5286 & 4.650 & 0.43 & 2.41 & 1.92 &     * \\
P\,64                 &  5267 & 4.618 & 0.95 & 2.81 & 3.32 &     * \\
P\,71$^{\mathrm{d}}$  &  5251 & 4.604 & 1.55 & 2.96 & 14.6 &    17 \\
P\,94                 &  5168 & 4.640 & 0.87 & 2.54 & 4.89 &    10 \\
P\,100$^{\mathrm{d}}$ &  4913 & 4.654 & 1.46 & 0.00 & 16.9 &    21 \\
P\,332                &  4989 & 4.660 & 0.76 & 0.00 & 2.34 &     * \\
P\,349                &  4917 & 4.662 & 0.57 & 0.00 & 3.89 &     * \\
P\,352                &  5105 & 4.658 & 0.87 & 0.00 & 4.30 &     * \\
 \noalign{\smallskip}
 \hline
 \end{tabular}
\begin{list}{}{}
\item[$^{\mathrm{a}}$] Calculated using Eq.\,(16.2) in Gray (1992); 
$^{\mathrm{b}}$ Calculated using the relation given in Fischer \& Valenti 
(2003) for stars of $T_{\rm eff}(B-V) > 5000$~K. For cooler stars, the values 
are set to zero; $^{\mathrm{c}}$ From Prosser \& Giampapa (1994); 
$^{\mathrm{d}}$ Variable star (c.f. The 73rd Name-list of Variable Stars, 
Kazarovets \& Samus 1997). 
\end{list}
\end{table*}

 \begin{table*}
 \tabcolsep 10pt
 \centering
 \caption{\label{abun} Element abundances on a logarithmic scale where H = 12}
 \begin{tabular}{lcccccccccc}
 \hline
 \hline
 \noalign{\smallskip}
 Star & Li & Li(MM)$^{\mathrm{a}}$ & O & Mg & Si & Ca & Ti & Cr & Fe & Ni\\
 \noalign{\smallskip}
 \hline
 \noalign{\smallskip}
Sun$^{\mathrm{b}}$&1.16&*&8.87&7.58&7.55 & 6.36 &5.02  &5.67 &7.50 & 6.25\\
P\,19 & 3.19&   *        &8.61&7.59&7.55& 6.24  &4.84  &5.82 &7.51 & 6.30\\
P\,147&2.64 &   *        &8.95&7.59&7.56 &6.35  &5.01  &5.68 &7.51 &6.24\\
P\,39 &3.02 &   *        &8.84&7.54&7.55 &6.40  &5.11  &5.61 &7.50 &6.17\\
P\,107&3.03 &3.0         &9.38&7.54&7.42 &6.36  &4.93  &5.59 &7.51 &6.16 \\
P\,150&3.19 &3.1         &9.22&7.91&7.32 &6.54  &5.23  &5.64 &7.46 &6.13\\
P\,151&3.01 &   *        &9.04&7.81&7.57 &6.49  &5.11  &5.58 &7.50 &6.16\\
P\,60 &3.02 &   *        &8.98&7.55&7.63 &6.42  &5.13  &5.69 &7.50 &6.19\\
P\,199& 1.51&   *        &9.09&7.57&7.49 &6.18  &5.56  &5.59 &7.43 &6.39\\
P\,75 & 3.33&3.3         &9.33&7.54&7.84 &6.51  &5.10  &5.50 &7.51 &  6.13\\
P\,165& 3.03&3.1         &9.10&7.42&7.39 &6.52  &5.27  &5.72 &7.47 &6.19\\
P\,267& *   &*           &8.84&7.57&7.61 &6.32  &5.38  &5.76 &7.42 &6.36\\
P\,64 &*    &*           &9.02&7.46&7.41 &6.17  &5.40  &5.82 &7.40 &6.37 \\
P\,71 & 2.96&3.1         &9.51&7.67&7.81 &6.52  &5.27  &5.48 &7.51 &6.17\\
P\,94 &1.44&2.1          &8.67&7.56&7.49 &6.50  &5.18  &5.60 &7.47 &6.18\\
P\,100& 2.66&2.8         &9.87&7.41&7.93 &6.48  &5.25  &5.54 &7.48 &6.27\\
P\,332&*    &*           &9.83&7.74&7.86 &6.14  &5.34  &5.56 &7.51 &6.59\\
P\,349&*    &*           &9.67&7.44&7.94 &6.17  &5.23  &5.57 &7.41 &6.50\\
P\,352&*    &*           &9.50&7.68&7.75 &6.23  &5.33  &5.48 &7.43 &6.48\\
 \noalign{\smallskip}
 \hline
 \end{tabular}
\begin{list}{}{}
\item[$^{\mathrm{a}}$] From Mart\'{i}n \& Montes (1997);  $^{\mathrm{b}}$ From Grevesse, Noels \& Sauval (1996)
\end{list}
 \end{table*}

\begin{figure}
\centering \epsfig{file=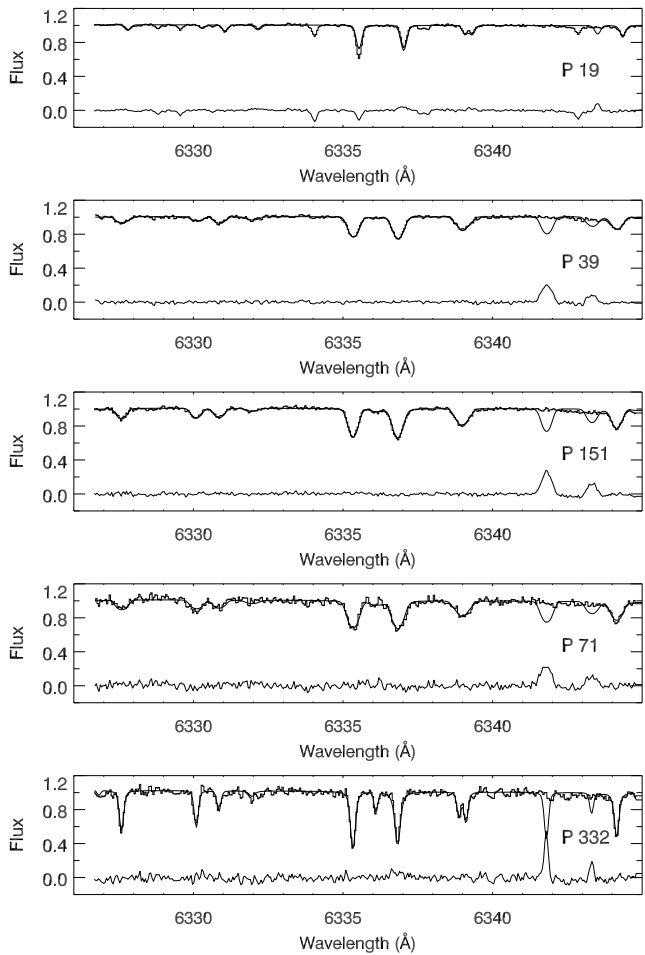, width=8.5cm, bbllx=74, bblly=3,
bburx=546, bbury=708, clip=, angle=0}
\caption{Sample spectra from 6326 to 6345\,\AA\ of five program stars,
selected to represent the different temperature regimes of the whole sample.
In each panel, the observed spectrum (histogram) is compared to the best-fit
synthetic spectrum (smooth curve). The residual spectrum of the fit is also
shown near the bottom of each panel. For all stars, the spectral region between
6341--6344\,\AA, as well as the \ion{Fe}{1} $\lambda$6335.33 line in P\,19
only, were excluded in the fitting.}
\label{specplot} \end{figure}

Note that because the strength of an absorption line depends not only on the
stellar surface chemical composition, but also on the excitation conditions as
well as on other stellar properties that affect the line profile, as one moves
from one star to another of a different spectral type, some lines strengthen
while others weaken, some new lines appear while some others disappear. Thus
the spectral line atomic data that we have determined by fitting the solar
spectrum, when applied to stars of other spectral types, may introduce some
errors in the results. However, for the limited range of spectral types covered
in the current study, this problem is not expected to be serious, especially
considering that we are more interested in the differential abundances, rather
than in the absolute elemental abundances.

Once the atomic data of spectral lines of interest have been determined by
fitting the solar spectrum, they are used to solve stellar parameters and
element abundances of target stars. In order to reduce the degeneracy amongst
the stellar global parameters, $\log\,g$ and macroturbulent velocity $v_{\rm
mac}$ were not optimized. Instead, we calculated $\log g$ using Eq.\,(16.2) of
Gray (1992) and estimated $v_{\rm mac}$ using the formula given in Fischer \&
Valenti (2003).  However, it is found that in some slow rotating late type
stars, adopting $v_{mac}$ thus obtained leads to negative values of $V\sin{i}$.
To avoid such unphysical situation happening, we arbitrarily set $v_{\rm mac}$
to zero for all stars of $T_{\rm eff}(B-V)$ lower than 5000\,K. 

A total of 33 stars were observed. Amongst them, one is later found to be a
double-line spectroscopic binary and two actually do not belong to IC\,4665.
In addition, 12 stars in the sample are found to have $B-V$ color temperatures
lower than 4800\,K. For these very cool stars, line blending becomes
increasingly problematic. Kurucz atmospheric models for such cool stars are
also known to be inadequate. We have therefore decided to leave out these stars
in our analysis. Our final sample contains 18 stars.  The global parameters of
these stars are presented in Table~\ref{para}.  Col.\,1 is the star name taken
from Prosser (1993). Cols.~2--3 give, respectively, $T_{\rm eff}$ determined
from our spectral analysis using SME and $\log\,g$ calculated from Eq.\,(16.2)
of Gray (1992).  Cols.~4--7 give, respectively, micro-turbulent velocities
determined from SME, macro-turbulent velocities calculated using the formula of
Fischer \& Valenti (2003) for stars of $T_{\rm eff} > 5000$~K, $V\sin{i}$
derived from SME, and finally $V\sin{i}$ given in Prosser \& Giampapa (1994).
Elemental abundances of Li, O, Mg, Si, Ca, Ti, Cr, Fe, and Ni derived from SME
are presented in Table~\ref{abun}.  Li abundances given in Mart\'{i}n \& Montes
(1997) are also listed in the Table for comparison. Oxygen abundances, deduced
from the \ion{O}{1} $\lambda$7774 triplet and discussed in details in a
separate paper (Shen et al. 2005; Paper~II thereafter), are included here for
completeness.

Sample spectra from 6326 to 6345\,\AA, one out of the seven spectral segments
chosen to determine global stellar parameters ($T_{\rm eff}$, $\log\,gf$, etc.)
are illustrated in  Figure~\ref{specplot} for five program stars, selected to
represent the different regimes of effective temperature spanned by stars of
the whole sample. The observed spectrum of each star (histogram) is compared to
the best-fit synthetic spectrum obtained with the SME. The residual of the fit
is also plotted near the bottom of each panel.  The region from 6341 to
6344\,\AA\, was masked out in the fitting. Also masked out was an unidentified
absorption feature near 6334\,\AA, observed only in the spectrum of P\,19.  In
general, the synthetic spectrum fits the observed line profiles well. An
exception was the \ion{Fe}{1} line at 6335.33\,\AA\, in the spectrum of P\,19.
The $\log\,gf$ value that we determined for this line from fitting the NSO
solar spectrum was $-2.289$, compared to the value of $-2.177$ retrieved from
the VALD.  Adopting the $\log\,gf$ value from the VALD did not improve the
fit, we eventually decided to exclude this line in the analysis of P\,19.

\section{Error analyses}

\subsection{Lunar corrections}

To validate and verify the accuracy of our SME solar solution, we have run the
SME on a lunar spectrum, applying the atomic data deduced from fitting the NSO
solar spectrum. The lunar spectrum was obtained by Dr. H.-W. Zhang using an
echelle spectrograph mounted on the National Astronomical Observatory 2.16\,m
telescope in Xinglong of China. The spectrum has a resolving power of $\sim
30,000$ and a S/N of approximately 300. The differences between the solar
parameters determined from the NSO solar spectrum and those determined from the
lunar spectrum are listed in Table~\ref{moon}. In all cases, the abundance
corrections are small compared to, for example, the spurious abundance
variations of yielded by the SME as reported and discussed in Valenti \&
Fischer (2005).  Given the small values of the lunar corrections and the
uncertainties in the applicability of those corrections to stars of spectral
type other than of the Sun, and considering that we are more interested in the
differential rather than the absolute abundances in the current study, we have
decided not to apply the lunar corrections to abundances deduced for our
IC\,4665 sample stars.

\begin{table}
\caption{\label{moon} Lunar corrections}
\centering
\begin{tabular}{lr}
\hline
\hline
\noalign{\smallskip}
Parameter & Correction~~\\
\noalign{\smallskip}
\hline
\noalign{\smallskip}
$T_{\rm eff}$& 28\,K\\
{[M/H]}& $-0.003$ \\
$v_{mic}$& $-0.05$\,km\,s$^{-1}$\\
O& $-0.03$ \\
Mg&0.03\\
Si&0.03\\
Ca&0.06\\
Ti&0.06\\
Cr&$-0.07$\\
Fe&0.02\\
Ni&0.02\\
\noalign{\smallskip}
\hline
\end{tabular}
\end{table}

\subsection{Effective temperature}

\begin{figure}
\centering
\epsfig{file=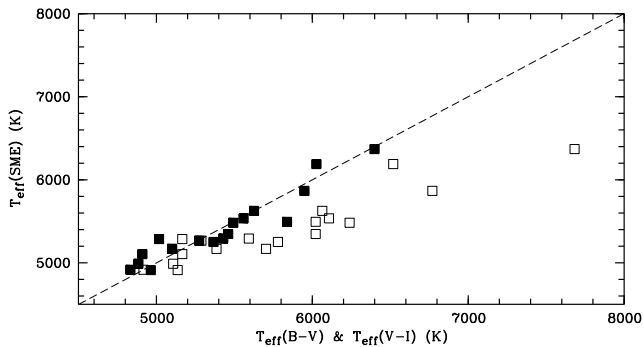,
width=8.5cm, bbllx=28, bblly=419, bburx=526, bbury=687, clip=, angle=0}
\caption{SME spectroscopic temperatures plotted against values derived
from the $(B-V)$ (filled squares) and from the $(V-I)$ color indices (open
squares).}
\label{teff}
\end{figure}



 \begin{table*}
 \tabcolsep 10pt
 \centering
 \caption{Comparison of effective temperatures} 
 \label{tefftab}
\begin{tabular}{lccccc}
 \hline
\hline
 \noalign{\smallskip}
 Star& $V$ & $B-V$ &$ V-I$ & $T_{\rm eff}(B-V)$ (K) & $T_{\rm eff}(V-I)$ (K) \\
 \noalign{\smallskip}
 \hline
 \noalign{\smallskip}
        P\,19  & 11.95 & 0.64 & 0.53 & 6399 & 7682  \\
        P\,147 & 13.45 & 0.73 & 0.74 & 6026 & 6519  \\        
        P\,39  & 12.93 & 0.75 & 0.69 & 5950 & 6770  \\
        P\,107 & 12.94 & 0.84 & 0.84 & 5626 & 6064  \\
        P\,150 & 13.08 & 0.86 & 0.83 & 5558 & 6107  \\
        P\,151 & 13.57 & 0.78 & 0.85 & 5837 & 6021  \\
        P\,60  & 13.43 & 0.88 & 0.80 & 5493 & 6239  \\
        P\,199 & 14.59 & 1.01 & 1.02 & 5100 & 5385  \\
        P\,75  & 13.70 & 0.89 & 0.85 & 5460 & 6021  \\
        P\,165 & 13.40 & 0.90 & 0.96 & 5428 & 5592  \\
        P\,267 & 14.83 & 1.04 & 1.09 & 5017 & 5166  \\
        P\,64  & 14.32 & 0.95 & 1.05 & 5274 & 5288  \\
        P\,71  & 13.68 & 0.92 & 0.91 & 5365 & 5779  \\
        P\,94  & 14.26 & 1.01 & 0.93 & 5100 & 5703  \\
        P\,100 & 14.37 & 1.06 & 1.10 & 4963 & 5136  \\
        P\,332 & 14.54 & 1.09 & 1.11 & 4884 & 5107  \\
        P\,349 & 14.65 & 1.11 & 1.18 & 4833 & 4916  \\
        P\,352 & 14.62 & 1.08 & 1.09 & 4910 & 5165  \\
\noalign{\smallskip}
 \hline
 \end{tabular}
 \label{linelist}
 \end{table*}

Table~\ref{tefftab} gives effective temperatures derived from the $B-V$ and
$V-I$ color indices. The color indices were taken from Prosser (1993). $T_{\rm
eff}(B-V)$ was derived using the formula of Alonso et al. (1996).  From
$E(B-V)=0.18$ (Mermilliod 1981a), we obtained a $V-I$ color excess of
$E(V-I)=0.23$ using the equation of Dean et al. (1978). Since the $I$
magnitudes were observed using the Kron system, we calculated the $V-I$ color
temperature using the equation given by Randich et al. (1996), 
\[
T_{\rm eff} = 9900-8598(V-I)_0+4246(V-I)_0^2-755(V-I)_0^3,
\]
where $(V-I)_0 = (V-I) - E(V-I)$.

In Figure~\ref{teff} we plot effective temperatures derived from the SME (c.f.
Table~\ref{para}) against those derived from the $(B-V)$ and from the $(V-I)$
color indices. Overall, color temperatures calculated from the $(B-V)$ color
index agree well with values determined using the spectroscopic method. For the
18 sample stars, differences between the two temperatures have an average value
of mere 2~K and a standard deviation of 141~K. Parts of the scatter are likely
caused by inhomogeneous reddening towards individual stars -- in our analysis,
we have adopted a constant reddening constant of $E(B-V)=0.18$ and
$E(V-I)=0.23$.  In order to examine the possible variations of reddening
towards the individual stars, we have measured the equivalent widths (EWs) of
the diffuse interstellar absorption band at 6613\,\AA. The values of
EW($\lambda$6613) are found to vary from 0.057 to 0.137\,\AA\, with typical
uncertainties of less than 0.005\,\AA. If we use the fitted linear relation
between $E(B-V)$ and EW(6613) given by Cox et al. (2005), a difference of
0.08\,\AA\, in EW($\lambda$6613) translates into a variation of $\sim$0.4 in
$E(B-V)$, and a corresponding variation of $\sim$1000~K in $T_{\rm eff}$. On
the other hand, we find that if we adopt the reddening implied by the measured
EW($\lambda$6613) for individual stars when calculating $T_{\rm eff}(B-V)$, the
resultant differences between $T_{\rm eff}(B-V)$ and $T_{\rm eff}$(SME) become
larger, not reduced, implying that inhomogeneous reddening is probably not the
main cause for the observed discrepancy between $T_{\rm eff}(B-V)$ and $T_{\rm
eff}$(SME).  Alternatively, given the large scatter in the relation
between the observed EW($\lambda$6613) and $E(B-V)$, it is possible that
the increased discrepancies between $T_{\rm eff}(B-V)$ and $T_{\rm eff}$(SME)
are caused by EW($\lambda$6613) not being an accurate enough indicator of
$E(B-V)$. In contrast to the $(B-V)$ color temperatures, Fig.\,\ref{teff}
shows that $(V-I)$ color temperatures are systematically higher than the SME
values by approximately a constant offset of about 411~K, clearly too large to
be accounted for by measurement uncertainties. Although $(V-I)$ color index is
generally believed to be a better temperature indicator than $(B-V)$, as the
former is less affected by uncertainties in reddening corrections, the close
agreement between $T_{\rm eff}(B-V)$ and $T_{\rm eff}$(SME) makes us to believe
that the $(V-I)$ colors for those IC\,4665 stars are possibly unreliable. 

Stars in young open clusters are known to be susceptible to strong surface
activities. In a separate paper devoted to oxygen abundance (Paper~II), we show
that there is strong evidence suggesting that stellar activities are
responsible for the anomalous oxygen abundances deduced from the \ion{O}{1}
triplet lines for dwarfs in IC\,4665 -- the abundances increase by almost an
order of magnitude as the effective temperature decreases from 6400 to 4900~K.
Similar trends have previously also been observed in two other young clusters,
the Pleiades and M\,34.  One of the major observational consequence of strong
surface activities is that they will generate color anomalies. As such, it is
possible that for young cluster stars, color indices are no longer good
diagnostics for stellar effective temperature.  It is suggested that high
surface magnetic fields in active stars can power enhanced chromospheric
emission and affect magnetically sensitive lines. And as a consequence,
non-thermal radiation related to stellar activity may lead to corrupted
temperatures estimated using the spectroscopic method (Gaidos \& Gonzalez
2002). On the other hand, no correlation is found between the abundance derived
from a particular line and its Land$\acute{\rm e}$ $g$ factor (Drake \& Smith
1993; Steenbock \& Holweger 1981) and between the temperature discrepancy and
indices of stellar activity (Gaidos \& Gonzalez 2002).  We believe that
spectroscopic effective temperatures derived in the current work using SME are
probably more reliable than those derived using photometric methods.

\subsection{Error budgets}

 \begin{table*}
 \tabcolsep 8pt
 \centering
 \caption{\label{err} Error budgets caused by uncertainties in $T_{\rm eff}$,
$\log\,g$, $v_{\rm mac}$ and those generated in the process of line profile fitting}
\begin{tabular}{lrrrrrrrrr}
 \hline
\hline
 \noalign{\smallskip}
 &Li~& O~&Mg~ &Si~&Ca~&Ti~&Cr~&Fe~&Ni~\\
 \noalign{\smallskip}
 \hline
\noalign{\smallskip}
\multicolumn{10}{c}{P\,19 ($T_{\rm eff}=6370$\,K)}\\
 \hline
 \noalign{\smallskip}
$T_{\rm eff}-100$~K     &$-$0.07  & 0.09 & $-$0.01 & $-$0.02 & $-$0.04 &  $-$0.04& $-$0.09 & $-$0.05  &$-$0.05 \\
$T_{\rm eff}+100$~K     &0.07 & $-$0.05 &  0.04 &  0.04&   0.07&  0.04&   0.01 &  0.06&   0.07\\
$\log\,g-0.1$           &0.00 & $-$0.03 &  0.02  & 0.01  & 0.02   &$-$0.02  &$-$0.03   &0.02   &0.01  \\
$\log\,g+0.1$           &0.00 &  0.05  &$-$0.01  & 0.01&  0.00 &   0.02 & $-$0.05 & $-$0.01   &0.01\\
$v_{\rm mic}-30$\%      &0.00&  0.00 &  0.01  & 0.01  & 0.01 &   0.00 & $-$0.07 & 0.00 &  0.01 \\
$v_{\rm mic}+30$\%      &0.00&  0.00 &  0.01&   0.00 & 0.00&   0.00 & $-$0.04  &0.00&   0.00\\
Total Err            &0.08 &0.10   &0.06& 0.05&    0.06&   0.06&  0.12 &0.07&0.08\\
\hline
\noalign{\smallskip}
\multicolumn{10}{c}{P\,151 ($T_{\rm eff}=5494$~K)}\\
 \hline
 \noalign{\smallskip}
$T_{\rm eff}-100$~K &  $-$0.12&   0.15&  $-$0.05 &  0.02&  $-$0.10  & $-$0.11&  $-$0.10&  $-$0.08 & $-$0.05 \\
$T_{\rm eff}+100$~K &  0.11 & $-$0.24 &  0.02 & $-$0.01  & 0.09   &  0.09   &0.10 &  0.06&   0.04  \\
$\log\,g-0.1$       &0.01 & $-$0.01 & $-$0.01 & $-$0.01  & 0.03 &   0.00 & $-$0.01 &  0.01 & $-$0.02\\
$\log\,g+0.1$       &  -0.01 &  0.01 & $-$0.01 & 0.00&  $-$0.04 &  $-$0.01 & 0.00 & $-$0.02  & 0.01  \\
$v_{\rm mic}-30$\%  & 0.02  & 0.03 & $-$0.04 &  0.00 &  0.14   & 0.07  & 0.03 &  0.07 &  0.02 \\
$v_{\rm mic}+30$\%  & 0.01  & $-$0.02&   0.05 & $-$0.01&  $-$0.16 & $-$0.08&  $-$0.02&  $-$0.08 & $-$0.03 \\
Total Err        & 0.13  &0.24 &   0.11    &0.06  &0.19 &0.14 &0.12       &0.11 &   0.11\\
\hline
 \noalign{\smallskip}
\multicolumn{10}{c}{P\,332 ($T_{\rm eff}=4989$~K)}\\
 \hline
 \noalign{\smallskip}
$T_{\rm eff}-100$~K &$-$0.15$^{\mathrm{a}}$   &0.12&  0.00 &  0.08  &$-$0.11   &$-$0.14  &$-$0.13&  $-$0.05&  $-$0.03\\
$T_{\rm eff}+100$~K & 0.13$^{\mathrm{a}}$&-0.15   &0.03&  $-$0.05&   0.10 &  0.12  & 0.10   &0.03 &  0.02 \\
$\log\,g-0.1$       &  0.01$^{\mathrm{a}}$ & $-$0.03  &$-$0.03  & 0.00  & 0.02   & 0.01  &0.00&  $-$0.02&  $-$0.03 \\
$\log\,g+0.1$       &$-$0.01$^{\mathrm{a}}$& 0.03   &0.02   &0.01 & $-$0.04   &$-$0.03 & $-$0.02  &$-$0.01&   0.01 \\
$v_{\rm mic}-30$\%  &0.01$^{\mathrm{a}}$&$-$0.01 & $-$0.01 & $-$0.02  & 0.04   & 0.04 &  0.01  &0.00&   0.01  \\
$v_{\rm mic}+30$\%  &$-$0.01$^{\mathrm{a}}$&$-$0.03  & 0.04  & 0.02 & $-$0.06   & $-$0.03  &$-$0.02  &$-$0.02 & $-$0.01 \\
Total Err        &0.19$^{\mathrm{a}}$  &                      0.19      &0.13   &0.15   &0.15    &0.16    &0.17    &0.06   &0.12\\       
\noalign{\smallskip}
 \hline
 \end{tabular}
\begin{list}{}{}
\item[$^{\mathrm{a}}$] For star P\,100 ($T_{\rm eff} = 4913$~K), as the Li line in the spectrum of P\,332 is too weak to measure.
\end{list}
 \label{linelist}
 \end{table*}

In order to obtain estimates of the possible errors of individual elemental
abundances derived using the SME, caused by uncertainties in the SME determined
stellar global parameters, including $T_{\rm eff}$, $\log\,g$ and $v_{\rm
mic}$, we have rerun the SME by varying $T_{\rm eff}$ by amounts of $\pm
100$\,K, $\log\,g$ by $\pm 0.1$ and $v_{\rm mic}$ by $\pm 30$\% from the
optimal values, for three stars, selected to represent different temperature
regimes spanned by the sample stars. The resultant changes in individual
elemental abundances are listed in Table~\ref{err}.  Errors generated in the
process of line profile fitting, caused by limited S/N ratios of the diagnostic
lines, were estimated by varying the elemental abundance and then analyzing its
effects on the residuals of the fit.
In the case of iron, as we used about 100 iron lines to determine its abundance 
and the lines distribute in all the seven spectral segments, 
we estimate an error from the scatter of abundance values deduced from the seven spectral segments individually. 
The total error budget, 
after adding these uncertainties in quadrature, are listed in Table~\ref{err}.

We show in the previous subsection that the differences between the SME
temperatures and the $(B-V)$ color temperatures have a standard deviation of
about 141~K. We believe that a larger part of the scatter is caused by
uncertainties in the color temperature rather than in the SME temperature, as
the former is probably affected by stellar surface activities which produce
color anomalies. Thus our assumption that the SME temperatures are probably
accurate to $\pm 100$~K is probably realistic. If we increase the uncertainties
of $T_{\rm eff}$ to $\pm 200$~K, the resultant errors of Fe abundances increase
to about 0.1--0.2\,dex.

\section{Abundance Dispersions}
\label{sec:disc1}

\begin{figure}
\centering
\epsfig{file=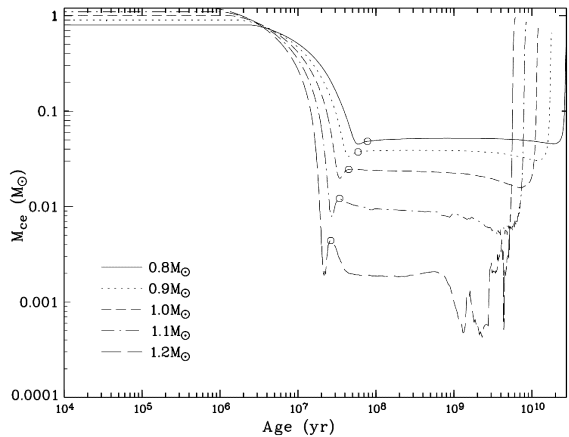,
width=8cm, bbllx=12, bblly=12, bburx=584, bbury=451, clip=, angle=0}
\caption{Evolution of CZ mass for stars of 0.8--1.2\,$M_{\odot}$. The
circle on each curve denotes the position of the zero age main sequence.}
\label{cz}
\end{figure}

In this section, we discuss the implications of metallicity dispersion on the
protracted accretion of protoplanetary material under the assumption that the
observed heavy elemental abundances only reflect those of the stars' surface
CZ.

\subsection{Iron abundance dispersion and CZ mass}

Because of their much lower masses, planets, including gas giants, contain
much smaller amounts of heavy elements than their host stars. The accretion of
planetary material can only make an observable difference if it contaminates
only the thin outer CZ of solar-type stars. In the scenario that SWPs acquire
higher than the average metallicity via accretion of metal-rich planetary
material, one expects that stars with a relatively shallow CZ will exhibit a
greater degree of pollution than those with a deep CZ.

The evolution of CZ mass ($M_{\rm CZ}$) with time for stars of mass $M_\ast$
between 0.8 and 1.2\,$M_\odot$, obtained using Eggleton's stellar evolution
code (Eggleton 1971; Pols et al. 1995), is plotted in Figure~\ref{cz}.  The
Figure shows that over time scales comparable to the age of the open cluster
IC\,4665 (30--40 Myr), $M_{\rm CZ}$ of young stars of $M_\ast > 0.8 M_\odot$
has already declined substantially and stabilized to its main-sequence value --
about 0.002\,$M_{\odot}$ for F dwarfs and $\ga 0.01$\,$M_{\odot}$ for G and
later type dwarfs. 

\begin{figure}
\centering
\epsfig{file=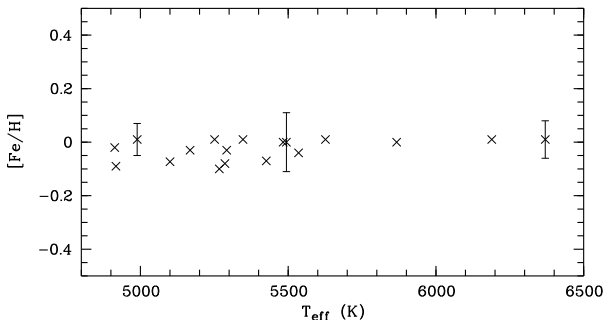,
width=8cm, bbllx=48, bblly=417, bburx=557, bbury=683, clip=, angle=0}
\caption{Iron abundance against $T_{\rm eff}$. Error bars for three selected
stars of different temperatures are also plotted.}
\label{fe}
\end{figure}

For main sequence stars, effective temperature increases with stellar mass.  
To search for possible correlation between metallicity and CZ mass, we plot in
Figure~\ref{fe} [Fe/H] against $T_{\rm eff}$ for all stars in our sample.
Within the observational uncertainties of individual stars, $\epsilon \sim
0.06$--0.11\,dex, no correlation is found.  A first order polynomial fit to the
data yields a slope of essentially zero ($\sim 10^{-5})$ and a linear
correlation coefficient of 0.27. In addition, the upper boundary of the
measured metallicities does not show any evidence of rising as $T_{\rm eff}$
increases (i.e. as CZ mass decreases), as one would expect in the accretion
scenario.  We note however that only a few stars in our sample have $T_{\rm
eff}$ higher than 5500\,K. The 18 stars in our sample yield an average iron
abundance of [Fe/H] = $-0.03$ and a standard deviation of $\sigma = 0.04\,$dex,
while the full range of [Fe/H] is about 0.15\,dex.  The measurement
uncertainties for hottest stars (i.e. most massive ones) in our sample are
about $\epsilon = 0.07$\,dex.

For an iron abundance of [Fe/H], the mass fraction of all refractory elements
is about $Z \simeq 0.003 \times 10^{[{\rm Fe/H} ]}$. If all heavy elements,
both refractory and volatile, are included, then the corresponding mass
fraction will be 6 times larger, i.e. $Z \simeq 0.018 \times 10^{[{\rm Fe/H}
]}$ (D$\ddot{\rm a}$ppen 2000). Thus an upper limit of enrichment $\delta$ by
accretion implies a maximum mass of accreted material which is given by,
\begin{equation}
M_{\rm acc} \la \delta  Z M_{\rm CZ}. \label{eq:dmacc} 
\end{equation} 

The hottest star in our sample is P\,19. Based on the Age-Zero models of
Drilling \& Landolt (2000), we estimate that P\,19 has a mass between
1.4--1.2\,$M_{\odot}$ and a CZ mass $\sim 0.002$\,$M_{\odot}$. The star has an
iron abundance of [Fe/H] = 0.01, which is $\Delta = 0.04\,$dex higher than the
cluster mean, and may have a metallicity that is at the maximum $\Delta +
\epsilon = 0.11$\,dex richer than the cluster mean. An enrichment at this
level, i.e. $\delta = 0.11\,{\rm dex} = 0.3$, caused by accretion of planetary
material, would thus require 0.6\,$M_\oplus$ of refractory material, including
iron. The limit would be raised to 3\,$M_\oplus$ if all volatile heavy elements
are augmented. These mass upper limits would be reduced to 0.3 and
2\,$M_\oplus$, respectively, if the constraint is applied with the measurement
uncertainty (i.e.  0.07\,dex), or to 0.2 and 1\,$M_\oplus$, respectively, if
the constraint is applied with the difference between the most likely value and
the cluster mean value (i.e. 0.04\,dex).

In the above discussion, we have adopted $\delta = \Delta + \epsilon$ to obtain
a stringent upper limit on the amount of pollution. The standard deviation of
iron abundance for the whole sample appears to be smaller than the measurement
uncertainty of individual stars (i.e. $\sigma < \epsilon$), suggesting that 1)
$\sigma$ is an upper limit for the metallicity dispersion $\delta$ and 2) our
estimate for $\epsilon$ may be over conservative. [In contrast, Wilden et al.
(2002) may have under estimated the magnitude of $\epsilon$.]  Based on the
assumption that the sixteen G-type stars in our sample provide an statistically
significant data set, we may set $\delta = \sigma \simeq 0.04\,{\rm dex} = 0.1$
in Eq.\,(\ref{eq:dmacc}).  For an $M_\ast = 1\,M_\odot$ G dwarf, $M_{\rm CZ} =
0.02\,M_\odot$. Therefore $M_{\rm acc} \la 2\,M_\oplus$ for refractory material
only. The constraint on $M_{\rm acc}$ when the volatile heavy elements are also
included is 10\,$M_\oplus$, which is comparable to that ($\sim
10$--40\,$M_\oplus$) inside Jupiter and Saturn (Guillot et al. 2004). For a
0.8$\,M_\odot$ star with $M_{\rm CZ} = 0.04\,M_\odot$, the lack of metallicity
dispersion implies $M_{\rm acc} \la 4\,M_\oplus$ in refractory and
20\,$M_\oplus$ in volatile heavy elements which are comparable to the total
mass in all the terrestrial planets and in a single gas giant, respectively.

\subsection{Lithium abundance}

Lithium abundances are listed in Table~\ref{abun} and compared to those
previously measured by Mart\'{i}n \& Montes (1997). The results agree well,
except for P\,94. Both analyses yield similar effective temperatures for this
star -- they find $T_{\rm eff} = 5135$\,K, compared to our value of 5168\,K.
The discrepancy is found to be caused by differences in the observed line
strengths. Our spectrum yields an equivalent width of 30\,m\AA\ for the
\ion{Li}{1} $\lambda$6708 resonance line, much smaller than the value of
90\,m\AA\ given in their paper.

It is well established that lithium burning during the pre-main-sequence
evolution leads to its depletion in young stellar clusters (Soderblom 1995).
Observation of lithium depletion in Hyades indicates that additional mixing
mechanisms must be at work apart from the standard convective mixing (e.g.
Schatzman \& Baglin 1991). The presence of additional mixing processes will not
only lead to the destruction of lithium in the stellar surface layer, but can
also reduce the level of enrichment (if any accretion of heavy elements occurs)
by efficiently flattening the radial abundance profile generated by the
accretion of planetary material and therefore hinder the detection of the
enhancement. For example, Vauclair (2004) has recently studied the thermohaline
convection induced by the inverse metallicity gradient and shown that if the
negative abundance gradient produced by the accretion of H-deficient material
exceeds a certain threshold, ``metallic fingers'' might be created that dilute
the accreted material inside the star.  On the other hand, since IC\,4665 is
much younger than Hyades ($\sim 800$\,Myr), any diffusion, if occurred, may not
have had time to dilute the polluted CZ of IC\,4665 stars.

\begin{figure}
\centering
\epsfig{file=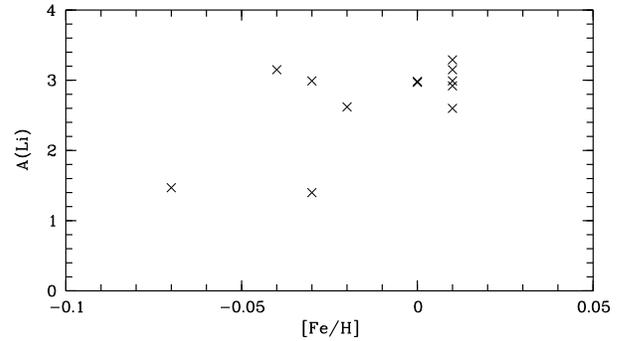,
width=8cm, bbllx=41, bblly=414, bburx=526, bbury=687, clip=, angle=0}
\caption{Li abundance against [Fe/H].}
\label{feli}
\end{figure}

Table~\ref{abun} indicates a factor of three spread in lithium abundance,
$A$(Li), amongst the IC\,4665 stars which have essentially identical iron
abundances. The large dispersion in the $A$(Li) suggests that the lithium
depletion time scale is comparable to the age of the cluster and therefore the
signature of planet consumption, if present, would be preserved. Following the
discussion in Section~1, we now consider the possibility that the extent to
which the large dispersion in $A$(Li) may be due to the late accretion of
lithium-rich protoplanetary material. If accretion of planetary material is
indeed responsible for the high lithium abundances observed in several of the
stars and for large scatter of $A$(Li) of the whole sample, then stars with
enhanced $A$(Li) should also exhibit enhanced abundances of other heavy
elements.  Alternatively, if the large dispersion in $A$(Li) is entirely caused
by physical processes other than accretion, one would expect no correlation
between the lithium abundance and those of other metals, and that all stars of
a given open cluster should have similar metal abundances.  To discriminate
these two possible scenarios, we plot in Figure~\ref{feli} lithium abundance
against that of iron for stars in our sample.  No obvious correlation is found.
However, there is some marginal evidence indicating that stars of higher
lithium abundances also have slightly higher iron abundances, up to [Fe/H] =
0.01, compared to the sample mean of [Fe/H] $ = -0.03$. This marginal iron
enhancement is however well within the measurement uncertainties.

Our results from the analysis of IC\,4665, i.e. the absence of any correlation
between [Fe/H] and $M_{\rm CZ}$ and between $A$(Li) and [Fe/H], are consistent
with the previous finding of Pinsonneault et al. (2001), Santos et al. (2004),
and Fischer \& Valenti (2005).  However, such a correlation (Laughlin \& Adams
1997) would be expected if the impinging planets are entirely disrupted in the
thin stellar CZ.  All these arguments suggest that pollution by accretion of
metal-rich planetary material is probably not the major contributor to the
observed high metallicity of SWPs.

\subsection{Other elements}

Gonzalez (1997) points out that since light elements are generally strongly
depleted in rocky planetesimals, the existence of a correlation between the
elemental abundances and their condensation temperatures could be a strong
signature of late-stage accretion of planetary material.  Tentative evidence
pointing to such systematic abundance variations as a function of condensation
temperature has indeed been found by Smith et al. (2001) in some SWPs.
Unfortunately, the observed trends can easily be confused with those resulted
from the Galactic chemical evolution.  In the current study, we have determined
abundances for Li, O, Mg, Si, Ca, Ti, Cr, Fe, and Ni and the results are listed
in Table~\ref{abun}. If accretion is at work as suggested by Gonzalez (1997),
then one would expect an enhancement in the abundances of refractory elements
(those with condensation temperatures near or above that of iron) are enhanced
relative to those of volatile elements. Elements analyzed in the current work
have condensation temperatures (Lodders 2003) of 1142~K (Li), 180~K (O), 1336~K
(Mg), 1529~K (Si), 1517~K (Ca), 1582~K (Ti), 1296~K (Cr), 1334~K (Fe) and
1353~K (Ni). It is unfortunate that the abundances of oxygen, the only volatile
element analyzed here, show a spurious trend with effective temperature. As we
will show in a separate paper (Paper~II), the anomalous oxygen abundances,
deduced from the triplet permitted lines, seem to be strongly affected by
stellar surface activities and therefore do not represent the true photospheric
values.

\begin{figure*}
\centering
\epsfig{file=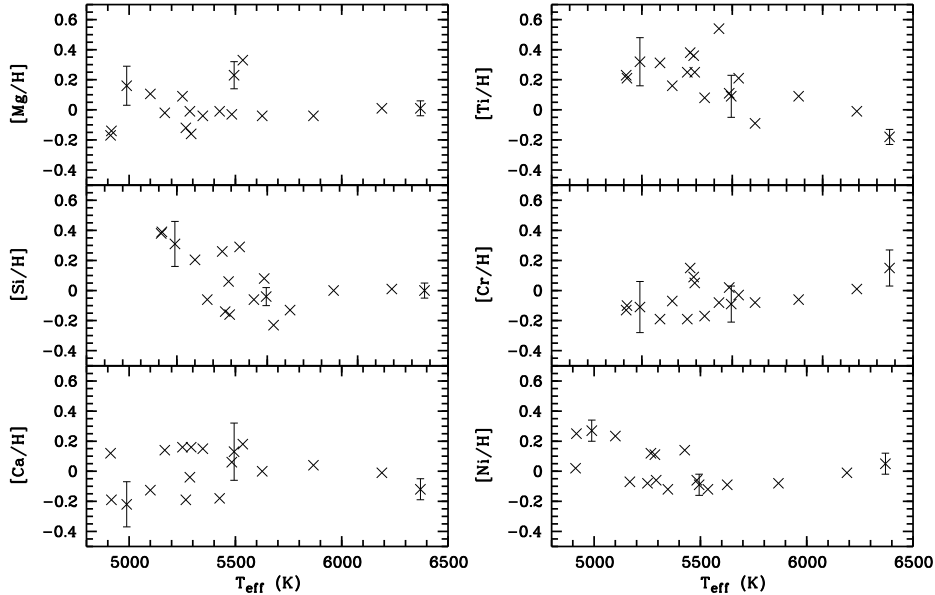,
height=8cm, bbllx=40, bblly=227, bburx=555, bbury=557, clip=, angle=0}
\caption{[X/H] as a function of $T_{\rm eff}$.}
\label{x}
\end{figure*}

\begin{table}
\caption{Average elemental abundances of IC\,4665 relative to the solar values}
\label{mean}
\centering
\begin{tabular}{cccccccc}
\hline
\hline
\noalign{\smallskip}
&[Mg/H]&[Si/H]&[Ca/H]&[Ti/H]&[Cr/H]&[Fe/H]&[Ni/H]\\
\noalign{\smallskip}
\hline
\noalign{\smallskip}
Mean&0.01&0.06&0.00&0.18&$-$0.05&$-$0.03&0.02\\
s.d.&0.13& 0.19&0.14&0.17&0.11&0.04&0.13\\
\noalign{\smallskip}
\hline
\end{tabular}
\end{table}

The abundances of Mg, Si, Ca, Ti, Cr and Ni are plotted in Figure~\ref{x}
against $T_{\rm eff}$. Except for Mg and Ca, all other four elements show
spurious trends with $T_{\rm eff}$. In the case of Si, Ti and Ni, the derived
abundance decreases with increasing $T_{\rm eff}$. For Cr, the trend is the
opposite. The average abundances of these elements as well as of iron are given
in Table~\ref{mean}, along with their standard deviations. Except for iron, the
abundances of all other elements show large scatters, with standard deviations
larger than 0.1\,dex, which are clearly too large to be accounted for by line
measurement uncertainties alone, and seem to be mainly caused by the spurious
variations with $T_{\rm eff}$ of the derived abundances.

Similar small but statistically significant variations with $T_{\rm eff}$ are
also observed in the Si, Ti, Cr and Ni abundances for stars of Pleiades
previously analyzed by Wilden et al. (2002) using the SME. In their study, the
abundances of Si and Ni are found to decrease by 0.1--0.2\,dex as $T_{\rm eff}$
increases from 4800 to 6000~K, whereas those of Ti and Cr increase by about
0.2\,dex. In their extensive study of over one thousand stars using the SME
Version\,2, Valenti \& Fischer (2005) found spurious trends in the derived
abundances of up to 0.3--0.4\,dex as $T_{\rm eff}$ varies from 4700 to 6200\,K.
They removed such trends by applying empirical corrections determined from
polynomial fits to the data. That elemental abundances determined for a
homogeneous sample such as an open cluster exhibit some unanticipated
correlation with stellar parameters such as $T_{\rm eff}$ is not a unique
feature of the SME, but is also found in other techniques such as those based
on equivalent width (EW) analysis. Using the EW method, Schuler et al. (2003)
determine elemental abundances for 9 solar-type dwarfs, spanning a temperature
range from 4700 to 6200\,K, in the open cluster M\,34 and find that cool stars
have 0.1--0.2\,dex lower Mg, Al, Ca, Cr, Ti and Fe abundances than hotter ones.
For Si the trend is reversed, whereas no significant trend is found for Ni.  In
their analysis, abundances of Si were derived from lines of relatively high
excitatio ($\chi\sim 6$\,eV), whereas for Fe, Ti, Cr, Ca, Al and Mg, most of
the lines used were of modest excitation ($\chi\sim 2$--4\,eV). Few Fe and Ni
lines in their analysis had excitation potentials higher than 5\,eV. Schuler et
al. suggest that NLTE effects are responsible for the observed spurious trends.
In our analysis of IC\,4665, Si abundances were derived from lines of $\chi\sim
5$--6\,eV, those of Cr from lines of $\chi\sim 0$--4\,eV and those of Ni from
lines of $\chi\sim 4$--5\,eV.  The observed spurious trends of abundances of
these elements with $T_{\rm eff}$ are therefore consistent with the picture of
Schuler et al.  However, in our analysis Ti abundances were derived from lines
of very low excitation ($\chi\sim 0$--2\,eV), yet they were found to decrease
with increasing $T_{\rm eff}$.  More observations and further quantitative
investigations of the NLTE effects are needed for a better understanding of the
underlying causes of these spurious abundance trends. 

\section{Metallicity Homogeneity in CZ's and the Planet Formation Processes}

In this Section, we discuss the implications of metallicity homogeneity among
the solar type stars in IC\,4665.  Similar to the previous Section, we assume
that the observed heavy elemental abundances only reflect those of the star's
surface CZ. 
Jeffery, Bailey and Chambers (1997) discussed the six possible
sources of material accreted onto the solar surface and estimated how much
rocky material of each source could have been accreted during the main-sequence
time of the Sun. In their analysis, protoplanetary disc debris and giant 
planet are the two main contributors that could give more than
1 $M_{\oplus}$ of infalling material to the solar surface. We will discuss the 
two sources one by one in this section. 
In Section~6.1 and Section~6.2, we will discuss the possibility of
grains or small planetesimals to be ejected into the CZ of its central star.
While in Section~6.3 and Section~6.4, the possibility of a giant planet to be
swallowed by its central star is discussed.

\subsection{Constraints on the evolution of debris disks}

The analysis in the previous Section suggests that it is unlikely that the
metallicity of the CZ of a SWP is significantly enhanced through post-formation
consumption of planets. Nevertheless, protracted accretion of protostellar
material is expected to proceed during the depletion of debris disks.  We now
utilize the above data to extract implications on the clearing of planetesimal
disks.

Based on the IR and millimeter excess continuum radiation, the total mass of
grains contained in a typical protostellar disk around T\,Tauri stars is
estimated to be in the range of $M_{\rm d} \sim 10^{-5}$--$10^{-2}$\,$M_\odot$
(Beckwith et al. 1990). In young clusters, the fraction of stars with
detectable traces of grain emission declines rapidly with the age on the time
scale of $\sim 3$\,Myr (Haisch et al. 2001; Carpenter et al. 2005).  The
reduction implies that grains are either accreted by their host stars or have
grown substantially. For stars in the T\,Tauri phase, since the CZ extends
throughout the whole star, $M_{\rm CZ}$ must be replaced by the total mass of
the star, $M_\ast$, when determining $M_{\rm acc}$ using Eq.(\ref{eq:dmacc}).
We will discuss the implication of abundance homogeneity using
Eq.(\ref{eq:dmacc}) in the next subsection.

Grain growth leads to the formation of planetesimals. During the post T\,Tauri
phase, planetesimals produce dust through collisions (Kenyon \& Bromley 2002).
Stars of ages comparable to that of IC\,4665 have been found to show a large
dispersion in their mid-IR excess, which is generally attributed to
reprocessed radiation emitted mostly by grains of sizes in the range of
10--100$\mu$m (Rieke et al. 2005). The observed mid-IR excess from an optically
thin circumstellar disk is given by,
\begin{eqnarray}
{F_{\nu }(\rm dust)\over {F_{\nu \ast}}} 
             & = & {{1\over D^2}\kappa _{\nu } 
	     B_{\nu }(T_{dust})M_{dust}}\over 
	     {{1\over D^2}\pi B_{\nu }(T_{\ast}){R_{\ast}}^2} \\
             & = & {1\over \pi }{T_{dust}\over T_{\ast}}
	     {1\over {R_\ast^2}}\kappa _\nu M_{dust},
\label{eqn:flux}
\end{eqnarray}
where $F_{\nu }(\rm dust)$ and $F_{\nu \ast}$ are flux densities from the dust
grains and from the central star, respectively, $D$ is the distance to the
source, $\kappa_\nu$ is the dust opacity, $T_\ast$ and $R_\ast$ are effective
temperature and radius of the central star, $T_{\rm dust}$ the average dust
temperature and $B_{\nu }$ the Planck function, approximated by the
Rayleigh-Jeans formula which is suitable for the mid- and far-infrared
wavelength regime of interest here.  

From the evolution of the 24\,$\mu$m excess deduced from observations by Rieke
et al. (2005), one finds that for a star of maximum mid-IR at age $t$, the
total mass of circumstellar dust grains is given by,
\begin{eqnarray} 
M_{\rm grain} & = & 2.22\times 10^{-8} [{150 ({\rm Myr})\over t}]
                    ({T_{\ast}\over T_{\odot}})({100\,{\rm K}\over T_{\rm dust}}) 
                    \nonumber \\
              &   & \times ({R_{\ast}\over R_{\odot}})^2 
                    ({0.02\,{\rm cm}^2\,{\rm g}^{-1}\over\kappa_{1.3{\rm mm}}})
                    ({\lambda_{\rm m} \over {1.3\,{\rm mm}}})~~ (M_\odot),
                    \label{eqn:grain} 
\end{eqnarray} 
where $\lambda_{\rm m}$ is the wavelength of maximum photon emission of the
grains, i.e. $T_{\rm dust}\lambda_{\rm m} = 0.367$\,cm\,K. If we assume an
opacity at 1.3\,mm, $\kappa_{\rm 1.3mm} = 0.5$\,cm$^2$\,g$^{-1}$ (for grains of
radii 0.1\,$\mu$m to 3\,mm, $\kappa_{\rm 1.3mm}$ varies in the range of 
0.14--0.87\,cm$^2$\,g$^{-1}$; Pollack et al. 1994) and a dust temperature of
150\,K (thus the grains have a maximum photon emission rate at 24\,$\mu$m),
then a solar type star at the age of IC\,4665 (i.e. $\sim 40$\,Myr) with the
maximum IR excess should have about $M_{\rm grain} = 4.1 \times
10^{-11}\,M_{\odot} = 1.3\times 10^{-5}\,M_{\oplus}$ dust grains.

The Poynting-Robertson (PR) drag by the central star's radiation causes the grains
to migrate toward the star on a time scale (Backman \& Paresce 1993),
\begin{equation}
\tau_{\rm PR} = 700 a_{\rm grain}\rho ({{r} \over {\rm 1 AU}})^2{L_{\odot }\over L_{\ast }}~~ ({\rm yr}).
\label{eqn:tau}
\end{equation}
For 24\,$\mu$m size grains of particle density $\rho = 3$\,g\,cm$^{-3}$ at a
distance $r$ of 1\,AU, $\tau_{\rm PR} = 5.0 \times 10^{4}$\,yr.  Under the
steady state assumption, the total amount of grains accreted over the age of
IC\,4665 is therefor $\sim (40\,{\rm Myr}/\tau_{\rm PR})M_{\rm grain} \simeq
0.01\,M_\oplus$. This value is much smaller than $M_{\rm acc}$ estimated in the
last Section from the observed metallicity dispersions. In other words, the
expected pollution due to the PR decay of the grains is well below the current
detection limit.

However, the grains have to be continually replenished by colliding
planetesimals. Under the assumption that the collisions between planetesimals
lead to an equilibrium power-law size distribution (Wetherill \& Stewart 1989),
\begin{equation}
dN/{d(a/ {a_0})} \simeq N_0({a/ {a_0}}) ^{-7/2},
\label{eqn:dn}
\end{equation}
and the collision frequency for planetesimals in the size range $a_{\rm p} \pm
\delta a_{\rm p}$ with all other smaller planetesimals is,
\begin{equation}
\omega_{\rm c} = A \sigma _{\rm p} n_s = \pi a_{\rm p}^2 \sigma _{\rm p} n_s,
\label{eqn:wc}
\end{equation}
where $A$ is the collision cross section, which is reduced to the geometrical
area for relatively high velocity dispersion $\sigma_p$ and $n_s$ denotes the
number density of planetesimals of sizes smaller than $a_p$. Assuming a
fraction $f$ of all collisions leads to planetesimals' total fragmentation and
the replenishment of planetesimals of smaller sizes, with a size distribution
as given by Eq.\,(\ref{eqn:dn}), an equilibrium state would be attainable.  We
assume the smallest size of planetesimals to be 24\,$\mu$m. The largest size
$a_{\rm max}$ of planetesimals, for which single collisions amongst them can
contribute to the replenishment of grains within the Poynting-Robertson
timescale $\tau _{PR}$, is estimated to be around $3\times 10^9$ cm. The
corresponding upper mass limit is 46 $M_{\oplus}$.  The actual upper limit of
the planetesimal size $a_{\rm upper}$ may be smaller than $a_{\rm max}$. Then,
the corresponding total mass of the parent-planetesimals becomes,
\begin{equation}
M_{\rm tot}\sim M_{\rm grain}({a_{\rm upper}\over a_{\rm min}})^{1/2}.
\label{eqn:12}
\end{equation}
Using $M_{\rm grain}$ calculated above, we obtain $M_{\rm tot} \sim 15
M_{\oplus}$, if $a_{\rm upper} = a_{\rm max}$.

Thus, the decline of $M_{\rm grain}$ with time also implies a diminishing
$M_{\rm tot}$.  In principle, depletion of the population of planetesimal
parent bodies can occur as a consequence of their captures by the host star.
This process is however unlikely the dominant outcome given that the inferred
reduction in $M_{\rm tot}$ is much larger than the upper limit $M_{\rm acc}$. A
more likely cause for the $M_{\rm tot}$ reduction is probably associated with
the accretion of planetesimals by relatively massive embryos and protoplanets.

\subsection{Constraints on terrestrial planet formation and late 
bombardment}

We now consider the constraints set by the upper limit of $M_{\rm acc}$ on the
extent of low-level metallicity pollution due to the protracted bombardment of
planetary material onto the host star.  With those constraints, we infer some
implications on the process of terrestrial planet formation.  

In the sequential accretion scenario, it is customary to adopt the minimum mass
nebula model as a fiducial prescription of initial condition for planet
formation in the solar system. The minimum mass nebula model is based on the
assumption that all the heavy elements in the primitive solar nebula have been
retained by the present-day terrestrial and giant planets. Current theories of
planet formation suggest that planetesimals grow through coagulation.  As a
consequence of their mutual gravitational scattering (Palmer et al. 1993) and
their tidal interaction with the disk gas (Artymowicz 1992), planetesimals
attain relatively modest eccentricities (Kominami \& Ida 2002).  With nearly
circular orbits, the growth of planetesimals is stalled when they evolve into
dynamically separated protoplanetary embryos with isolation masses (Lissauer
1993) which increases from a few times that of the Moon at 0.5 AU to a fraction
of Mars just inside the snow line (Kokubo \& Ida 2002).

Analogous to the decline of IR continuum radiation associated with hot, warm,
and cold dusts, the signature of gas accretion onto protostellar disks appears
to decrease on the time scale of 3--10 Myr (Hartmann 1998). There are also
evidence that the gas surface density in the inner disk also vanishes on a
comparable time scale (Najita 2003).  If gas in the outer regions of disks
decreases on a similar time scale, gas giant formation must proceed within a
few Myr. In contrast, the final assemble of terrestrial planets may occur on a
much longer time scale even though the initial growth of the grains may have
occurred during the first few Myr.  The giant-impact scenario for the origin of
the Moon suggests that collisions between protoplanets may have occurred after
they have differentiated (Cameron \& Benz 1991).  Independent cosmochemical
analysis based on the Hafnium/Tungsten isotopic abundances in meteorites and
the Earth lithosphere suggests that the final assemble of terrestrial planet
formation occurred on a time scale of 30--50 Myr in the Solar System (Yin et al. 2002).  Since this time scale is comparable to the age of IC\,4665 and that
for the CZ of solar-type stars to evolve to their asymptotic masses, we can
apply our data to extract useful implications on the formation of the
terrestrial planets.

In disks with heavy-element contents less than that of the minimum mass nebula,
gas giant cannot emerge prior to gas depletion (Ida \& Lin 2004a).
Nevertheless, embryos with relatively low isolation masses can emerge in the
inner regions of the disk.  After the gas depletion, these embryos'
eccentricity would be excited in a gas-free environment through their distant
gravitational interaction with each other (Chambers, Wetherill \& Boss 1996).
When their orbits cross with each other, their growth through coagulation would
resume. On the time scale of 100 Myr, this sequence of events leads to the
emergence of a small number of protoplanets with masses comparable to that of
the Earth and eccentricities comparable to the ratio of their surface escape
and orbital speed which is $\sim 0.1$--0.3 (Chambers \& Wetherill 1998).  In
contrast, the present-day eccentricities of the terrestrial planets in the
solar system are well below these values.

There are several scenarios to account for this discrepancy. Although
tidal interaction between planetesimals and the residual gas can
suppress the embryos' eccentricity, very delicate timing is required
for the emergence of sufficiently massive protoplanets and
sufficiently small eccentricities (Kominami \& Ida 2004).  Another
potential damping mechanism is dynamical friction exerted on the
embryos by low-mass planetesimals (e.g. Palmer
et al. 1993).  However, a population of
low-mass planetesimals with total mass comparable to that of the
terrestrial planets is needed to effectively reduce the eccentricity
of the latter (Goldreich et al. 2004). Some planetesimals may be
accreted by the terrestrial planets while a fraction of them may be
scattered into their host stars on a time scale comparable to the age
of IC\,4665.  The above constraint on the dispersion in [Fe/H] places an
upper limit on $M_{\rm acc}$ to be $2 M_\oplus$ for
$1 M_\odot$ stars.  

Other scenarios for terrestrial planets' low eccentricities rely on the
presence of gas giant planets.  In disks with masses comparable to that of the
minimum mass nebula model, gas giant planets can form just beyond the snow line
prior to the gas depletion.  In this limit, the giant planets' gravitational
perturbation can induce secular resonances which sweep over extended regions of
the disks during the gas depletion (Ward 1981). The eccentricity of
planetesimals along the path of the sweeping secular resonance is excited by
the gas giants while it is also damped by their tidal interaction with the
residual gas.  The combined influences of these two effects induce the
planetesimals to cross each other's orbits with aligned longitudes of periapse
and promote their growth through cohesive collisions (Nagasawa, Lin \& Thommes
2005).  After their masses become comparable to that of the Earth, the
protoplanetary cores become detached, with low-eccentricity orbits, from the
sweeping secular resonance of gas giants (Thommes, Nagasawa \& Lin 2005).  The
high retention efficiency of refractory planetesimals by the terrestrial
planets implies very limited amount of stellar pollution by protracted
planetesimal accretion, which is in good agreement with the metallicity
homogeneity we have established here.

The dynamical evolution from embryos to protoplanets is determined by the gas
depletion time scale as well as the gas giants' mass and eccentricity.  If
Jupiter and Saturn formed with their present eccentricities and gas in the
solar nebula was depleted over an e-folding time scale of 3\,Myr, the residual
planetesimals would be efficiently retained and assembled into terrestrial
planets with present-day masses and eccentricities over $\sim 20$--50 Myr (Lin,
Nagasawa, \& Thommes 2005).  But, if Jupiter and Saturn had much smaller
initial eccentricities or if the total mass of the gas in the solar nebula
rapidly declined below the mass of Jupiter, the gas giants would impose a weak
secular perturbation on the dynamical evolution of the planetesimals during the
propagation of their secular resonance.  When the mass of the disk gas
decreases below to that of the gas giants, their secular resonances are stalled
at asymptotic locations which are determined by either the dynamical
configuration of multiple gas giant planets or the relativistic precession in
systems with only single gas giant planets.  In this gas-depleted background,
the massive embryos scatter and accrete the low-mass residual planetestimals
and evolve into proto terrestrial planets with low eccentricities (Chambers \&
Cassen 2002).  A population of planetesimals are scattered onto the gas-giants'
secular resonances where they lose angular momentum to the gas giants, gain
eccentricity, and eventually collide with their host star. Possible
consequences of this process include 1) the intense cometary bombardment onto
$\beta$ Pic (Levison, Duncan \& Wetherill 1994) and 2) the large dispersion in
the IR excess of debris disks around young (Rieke et al. 2005) and mature stars
(Beichman et al. 2005). However, the observationally inferred upper limit on
$M_{\rm acc}$ suggests that the amount of planetesimals lost to their host
stars may not be sufficient to absorb all the excess kinetic energy that needs
to be lost during the circularization of the terrestrial planets.

\subsection{Implication on the evolution of short-period gas giants}

Although the scenario of gross metallicity enhancement in SWP's CZ
through planetary consumption is disproved by the constraint placed by
Eq.(\ref{eq:dmacc}), the possibility of one or two gas giants being
accreted by their host stars still cannot be ruled out with the 
present sets of data.  We discuss here some implications of this 
possibility.

Beyond the snow line, the condensation of volatile ices greatly increases the
isolation mass to several times that of the Earth.  Provided there is an
adequate reservoir (i.e. in relatively massive disks), gas is readily accreted
onto the cores initially through a phase of slow Kelvin-Helmholtz contraction
and followed by a phase run-away dynamical accretion.  Gas giant planets form
readily near the snow line during the active phase of protostellar-disk
accretion (Pollack et al. 1996).  When the gas giant planets acquire masses
comparable to that of Jupiter, these gas giants induce the formation of gaps in
the disk which not only quenches their growth, but also causes their orbits to
migrate with the viscous evolution of the disk (c.f. Lin \& Papaloizou
1986a,b).  This scenario has been invoked (Lin, Bodenheimer \& Richardson 1996)
to account for the origin of the first short-period planet discovered around a
main sequence star, 51 Peg (Mayor \& Queloz 1995).  Along their orbital decay,
gas giant resonantly capture and drive the residual planetesimals to migrate
with them.  As their eccentricity grows, these planetesimals may be scattered
into their host stars (Yu \& Tremaine 2001).  This process occurs while the
host stars are sufficiently young that their interior is fully convective.
Therefore it cannot significantly modify the metallicity of the host stars.

There are several other versions of the migration scenario.  Before the
protoplanets have sufficient mass to open up gaps in their nascent disks, they
tidally interact with both the disk region interior and exterior to their orbit
(Goldreich \& Tremaine 1980). Due to a geometric offset, it has been suggested
that Earth mass planetesimals may undergo Type-I migration (Ward 1986), migrate
to the stellar proximity, and form short-period gas giants in situ (Ward 1997).
It has also been suggested that the migration of gas giants may be induced by
their gravitational scattering of a large population of residual planetesimals
(Murray et al. 1998). The apparent chemical homogeneity of the stars in
IC\,4665 implies that if these processes occurred, the residual planetesimals
near the stellar surface must be cleared before the CZ significantly reduces
its mass.  Since the total mass of planetesimals needed is comparable to that
of the gas giants and the expected migration time scale is expected to be much
longer than 10--30 Myr, the planetesimals scattering migration, if occurred,
would introduce a metallicity enhancement much larger than that constrained by
our data.

In principle, the above constraint can be disregarded on the ground that only
1--2\% solar type stars host short-period gas giants. However, the period
distribution of extra solar planets is approximately flat over the range of
several days to months (e.g. Udry, Mayor \& Santos 2003).  Numerical
simulations of gas giant formation and migration reproduce the period
distribution in the weeks to years range but they also over predict the
frequency of short-period planets by nearly an order of magnitude (Ida \& Lin
2004b).  One possible resolution for this discrepancy is that a significant
fraction of the planets migrated to the disk centers either plunge into the
host stars (Sandquist et al. 1998) or are tidally disrupted (Gu et al. 2003,
2004).  These events may occur during the post-formation tidal evolution or
during the epoch of disk depletion in multiple- planet systems (Nagasawa \& Lin
2005).

In addition to tidal disruption, short-period gas giants are also vulnerable to
photo-evaporation and magnetic stripping through unipolar induction.  Outflow
is observed around HD\,209458b (Vidal-Madjar et al. 2003).  Although the mass
loss rate around its mature main sequence star is low, it could be considerably
higher during the early active phase of protostellar and planetary evolution.
Catastrophic mass loss from gas giants has been suggested as a potential avenue
for the formation of Neptune-mass planets (e.g. Boss et al. 2002).  If the
removed gas from the metal-enriched planetary envelopes is accreted by the host
stars, their metallicity would also be enhanced. The absence of any significant
metallicity dispersion implies that the debris of tidally-disrupted or
photo-evaporated gas giants is either blown away from host stars or accreted by
them before their CZ have established their asymptotic structure.  Based on
this result, we predict that the frequency of short-period planets around post
T Tauri stars is comparable to that of the mature stars.

\subsection{Implication on the evolution of eccentric gas giants
and multiple-planet systems}

The number of stars we have analyzed and presented for IC\,4661 and the
Pleiades clusters is small. The statistical significance of our constraints on
the short-period planets may be limited since they are found only in 1-2\% of
the target stars.  However, at least 10\% of nearby solar-type stars appear to
bear longer-period planets around them and in most cases these planets are in
multiple systems (e.g. Marcy et al. 2000a).  As the time baseline of precision
radial velocity data expand, the fraction of stars with planets is expected to
increase substantially (Armitage et al. 2002). Thus, the observed metallicity
homogeneity among the stars in IC\,4665, when combined with that obtained early
for the Pleiades stars (Wilden et al. 2002), does provide statistical
meaningful constraints on the pollution of planet bearing stars in general.

The eccentricity of the extra solar planets is nearly uniform between 0 and 0.7
(Marcy et al. 2000b). Although planets' resonant interaction with the disk has
been invoked as a potential mechanism for exciting their eccentricities
(Goldreich \& Sari 2003), both resonant and nonlinear damping processes limit
the extent of eccentricity excitation, especially for planets with mass less
than ten times that of Jupiter (Goldreich \& Tremaine 1980, Artymowicz 1992,
Papaloizou et al. 2001). 

An alternative scenario for the origin of large planetary eccentricities is
dynamical instabilities in multiple-planet systems. Such systems are expected
to be around most SWP's (Marcy et al. 2000a) because the emergence of the first
gas giants promotes the formation of the next-generation planets just beyond
the outer edge of the disk gap they induce (Bryden et al.  2000a).  Dynamical
instabilities can arise prior to (Bryden et al. 2000b; Kley 2003), during
(Nagasawa et al. 2003), and after disk depletion (Rasio \& Ford 1996;
Weidenschilling \& Marzari 1996; Lin \& Ida 1997). 

If SWP's were mostly formed in clusters, similar to most young stellar objects
(Lada \& Alves 2004), their encounters with each other could also perturb the
orbits of long-period planets and trigger dynamical instabilities in
multiple-planet systems (Laughlin \& Adams 1998). Finally, many planets,
including those with short periods, are found in binary stars.  If the
inclination between the stellar and planetary orbits are sufficiently large,
the Kozai effect may also induce planets to attain nearly parabolic orbits and
strike their host stars (Innanen et al. 1998; Murray et al. 1998) or trigger
dynamical instability in multiple systems.

Planets in dynamically unstable systems undergo orbit crossings and close
encounters which lead to large eccentricities, merges, and escape from their
host stars (Lin \& Ida 1997; Papaloizou \& Terquem 2001).  Under some
circumstances, planets may be scattered to the stellar proximity (Ford et al.
2005) or directly into the star.  Numerical simulations of planetary impacts
onto their host stars (Sandquist et al. 1998) show that a Jupiter-mass planet
cannot penetrate through the CZ of G dwarfs without being completely
disintegrated by the hydrodynamic drag.  But with the constraints set by
Eq.\,(\ref{eq:dmacc}), the observed metallicity homogeneity amongst the limited
number of stars we have analyzed is insufficient to rule out this possibility
for stars with masses $<1 M_\odot$.  This constraint is much more stringent for
F stars.  However, only a fraction of a one-Jupiter-mass impactor may be
removed from the planet as it penetrates directly through the much shallower CZ
of a $1.22\,M_{\odot}$ star. Depending on the actual location of most of the
heavy elements within the planet (possibly mostly confined in the core) and its
orbital impact parameters, it is not clear whether the accretion of a planet
onto an F-type star can result in greater metallicity enhancement of its CZ
than a G-type star.  With only one F star in our data set, we need further
quantitative studies of these processes to address the apparent discrepancy
between the expectations of the accretion scenario and observations such as
those presented in the current work.

\section{Constraints on the protoplanetary-disk masses} 

In this Section, we consider the alternative possibility that abundance is
uniform throughout the stars and use the metallicity dispersion to derive
constraints on the efficiency of planet formation in protostellar disks.  Based
on the analysis in \S~\ref{sec:disc1}, we adopt the standard deviation of
[Fe/H] ($\sigma \sim 0.04\,{\rm dex} = 0.1$) as the upper limit for the
metallicity dispersion.  

\subsection{Mixing in the protocluster cloud}

The remarkable abundance homogeneity among the stars in both IC\,4665 and the
Pleiades clusters cannot be attained unless their progenitor clouds are
thoroughly mixed.  These clouds are assembled from smaller clouds with
presumably a range of metallicity.  Turbulence provides a support against the
clouds' own self gravity as well as induces mixing to homogenize the clouds.
At least several eddie turn-over time scales are needed for the metallicity
dispersion to be reduced below the observed upper limit (Klessen \& Lin 2003). Thus, the progenitor clouds must be bona fide entities rather than
transitory flow pattern in the interstellar medium.

The total mass of these clusters is only a few $10^3\,M_\odot$. If the original
clouds have comparable masses, the injection of supernova remnants of one or
two O stars would be sufficient to significantly modify their metallicity.  The
absence of metallicity dispersion amongst the cluster stars suggests that the
span of star formation epoch in these clouds must have been shorter than the
main sequence life time of the massive stars which is $\sim 3$\,Myr.  This
inference is consistent with the range of stellar ages in the Taurus complex
(Cohen \& Kuhi 1978).  By examining the heavy element abundance patterns
of two Hyades candidates and seven members of the Ursa Major group, Gaidos \&
Gonzalez (2002) also reach a similar conclusion that any variation in
metallicity within a cluster is likely caused by heterogeneous incorporation of
heavy elements into protostars, rather than by the influence of massive
stars of a previous generation. 

\subsection{Mass of protostellar disks}

According to the current paradigm for star formation, all the gas in young
stellar objects has been processed through protostellar disks (Shu et al.
1993).  It is customary to classify protostellar evolution in four stages: 1)
The embedded phase extending over $< 0.1$\,Myr, during which massive, self
gravitating disks are embedded in collapsing progenitor clouds; 2) The active
disk evolution phase lasting $\sim 0.1$\,Myr, during which the energy
dissipation associated with mass diffusion and angular momentum transfer
contributes to most of the disk luminosity; 3) The passive disk evolution phase
proceeding over several Myr during which the disk mass is reduced to around
that of the minimum mass nebula and the disk's luminosity is mostly due to the
reprocessed stellar radiation; 4) The weak-line T Tauri phase evolving on the
time scale of several more Myr during which gas accretion onto the host stars
vanishes and the signature of the $\mu$m--mm size grains declines.

During each of these stages, gas and dust evolve independently.  In
differentially rotating accretion disks, the rate of gas diffusion is
determined by the efficiency of its angular momentum transport processes (Lin
\& Papaloizou 1996).  Disk gas may also be photo-evaporated at rates which are
determined by the intensity of the UV photons which can reach the outer disk
regions (Hollenbach \& Adams 2004).

In contrast, only sub-mm grains are well coupled and evolves with the disk gas.
The migration of larger-than-mm grains is determined by the effectiveness of
the hydrodynamic drag process (Takeuchi et al. 2005). Heavy elements contained
in super-km planetesimals are no longer significantly affected by the disk gas
(Garaud et al. 2004) and most of them are either retained by the protoplanets
or scattered beyond the orbits of the outermost planets.

The stars' asymptotic metallicity is determined not only by that of their
progenitor clouds, but also by the efficiency of dust retention and gas outflow
from the disk.  Due to the diverse avenues of evolution, the accretion of gas
and of heavy elements are not expected to proceed at the same rates.  Even if
the progenitor cloud of IC\,4665 is well mixed initially, different physical
processes could lead to an observable dispersion in the IC\,4665 stars.

There are two possible implications to this dichotomy: 1) the accretion of
condensible heavy elements is self regulated or 2) the retention efficiency of
heavy elements by protostellar disks is limited. The presence of planets in the
solar system implies a minimum mass nebula is the lower limit on the amount of
heavy elements which was left behind in the disk as planet-building blocks.
Current sequential planet models suggest that similar disks must be preserved
around a significant fraction of all young stellar objects in order to account
for the ubiquity of Jupiter-mass planets around nearby stars (Ida \& Lin
2004a,b).  At least during the passive phase of their evolution, protostellar
disks are observed with total grain masses comparable to that contained in the
minimum mass nebula (Beckwith \& Sargent 1991).

But, the orbital decay time scale for these grains $\tau_a$ is much less than
the disk evolution time scale and their retention requires their growth time
scale $\tau_g$ to be shorter than $\tau_a$.  In principle, dust can settle to
the mid-plane and become gravitationally unstable in turbulent-free
protostellar disks (Goldreich \& Ward 1973).  But the sedimentation also leads
to a strong shear layer (Weidenschilling \& Cuzzi 1993) which becomes unstable
well before the onset of gravitational instability (Sekiya \& Ishitsu 2000;
Garaud \& Lin 2004).  Gravitational instability can still occur in regions
where the abundance of heavy elements is comparable to that of the hydrogen and
helium (Sekiya 1998; Youdin \& Shu 2002). With such a large concentration of
heavy elements, cohesive collisions can also lead to $\tau_g < \tau_a$.  Such
high concentration of heavy elements may be accomplished through 1) sublimation
near the snow line (Stevenson \& Lunine 1988), 2) upstream diffusion (Morfill
\& Voelk 1984; Clarke \& Pringle 1988), 3) grain recycle through stellar winds (Shang et al. 2000),
4) differential grain migration (Youdin \& Chiang 2004), or 5) grain trapping
by persistent giant vortices (Adams \& Watkins 1995; Godon \& Livio 1999).  

In principle, all of these processes can occur at each stage of protostellar
evolution.  However, any one of these processes can also lead to the retention
of a large fraction of the heavy elements pass through the disk and introduce a
metallicity dispersion in their host stars.  Similar to the condensed heavy
elements, a fraction of disk gas may be evaporated either through
photo-evaporation (e.g. Hollenback \& Adams 2004) or winds driven by the
central stars.  Unless gas loss is coordinated with the retention of heavy
elements by planetesimals, these competing processes are likely diversify the
rates of gas and of heavy element accretion and generate metallicity dispersion
amongst the host stars.  Note that the dust migration rate actually increases
during the depletion of the disk as the hydrodynamic drag on the radial
migration of the grains is reduced (Takeuchi et al. 2005). A theoretical
challenge is to account for both the chemical homogeneity and the ubiquity of
planets.

The growth rate and isolation mass of planetesimals increase with $\Sigma_{\rm
d}$ and $\Sigma_{\rm d}^{3/2}$, respectively. In the limit of efficient grain
retentions, the surface density of planet-building blocks $\Sigma_{\rm d}$ is
much higher than that of the minimum mass nebula, which promotes the rapid
emergence of cores and gas giants (Ida \& Lin 2004a), even during the active
phases of disk evolution when the accretion rate onto their host stars is
greater than $10^{-7} M_\odot$ yr$^{-1}$.  After the first-generation planets
acquire sufficient mass to open gaps in their nascent disk, they migrate with
the disk gas.  Along their inwardly migrating path, these planets capture any
residual planetesimals and grains which are coupled to the gas onto their main
motion resonances and sweep them into their host stars (Ida \& Lin 2004b).
This self regulated clearing of the heavy elements would occur continually
until the disk is so depleted that the formation time scale of the last gas
giants becomes comparable to their orbital migration time scale, which is
determined by the mass diffusion and angular momentum transport time scale in
the disk.  The final stalling condition is satisfied when the mass of the
residual planetesimal disks is reduced to that of the minimum mass nebula.

The self-regulated disk-clearing scenario provides an attractive hypothesis to
resolve the [Fe/H]-homogeneity versus ubiquitous-planet paradox. But it also
predicts the existence of planets around a relatively large fraction of stars,
including metal deficient ones. This extrapolation is inconsistent with the
fact that the fraction of SWP's is observed to be a rapidly increasing function
of their metallicity (Santos et al. 2004; Fischer \& Valenti 2005).  The
observed correlation can be best reproduced by the sequential planet formation
models which are based on the assumption that the ratio ($\eta$) of the
retained to the total heavy elements accreted through the protostellar disks is
independent of the metallicity of their host stars (Ida \& Lin 2004b).  The
fiducial value of this ratio for the minimum mass nebula is $\eta > 0.02$
(Hayashi et al. 1985). The homogeneity of different mass stars indicates an
upper limit for this ratio to be $\eta < \sigma \sim 0.1$, which also appears
to be independent of $M_\ast$. (The constraints set by the stars in the
Pleiades cluster is $\eta < 0.05$; Wilden et al. 2002).  This upper limit means
that the total reservoir of heavy elements retained by their nascent disks is
less than five times that of the minimum mass nebula.  The cause for this
magnitude of $\eta$ is unclear.

The upper limit on $\eta$ also provides a constraint on the gas giant
formation.  In the early version of gas giant formation model,
10--20\,$M_\oplus$ cores are needed to initiate efficient gas accretion
(Pollack et al. 1996).  Such massive cores would emerge within the observed
disk depletion time scale of 3--10\,Myr if $\Sigma_{\rm d}$ is several times
larger than that of the minimum mass nebula.  The upper limit of $\eta$
determined here places a constraint on $\Sigma_{\rm d}$.  Modest values of
$\eta$ can still lead to the ubiquitous production of gas giant planets if 1)
$\Sigma_{\rm d}$ is enhanced near the snow line or 2) efficient gas accretion
is initiated with modest core masses.  The former possibility is consistent
with the present-day location of Jupiter being close to the snow line in the
solar nebula whereas the latter possibility is in agreement with the recently
revised upper limit on the core mass of Jupiter (Guillot et al. 2004).  The gas
accretion barrier may be bypassed with either an opacity reduction through
grain depletion (Ikoma, Nakazawa \& Emori 2000) or other efficient energy
transfer mechanisms though the radiative regions of the envelope.

\section{Summary}

In the current paper, we present the first abundance study other than Li for
the young open cluster IC\,4665.  Elemental abundances of Li, O, Mg, Si, Ca,
Ti, Cr, Fe and Ni have been determined for 18 dwarfs of spectral types from F
to early K, using high-resolution spectra obtained with the HiRes spectrograph
mounted on the Keck\,I 10\,m telescope.  Except for iron, abundances of all
other elements, including O, Si, Ti and Cr, show large scatters and trends with
effective temperature.  Similar trends are also observed in Pleiades (Wilden et
al. 2002), M\,34 (Schuler et al. 2003) and in field stars (Valenti \& Fischer
2005).  Except for Li, for which the abundance scatter is probably caused by
depletion, variations observed in other elements are probably spurious,
although their causes remain unclear.  In the case of oxygen, as we will show
in a separate paper (Paper~II), there is strong evidence suggesting that
stellar surface activities are to blame for the observed large variations of
abundances deduced from the $\lambda\lambda$7772,7774,7775 triplet lines.
Accurate abundance determinations for open clusters are pivotal to constrain
the formation and evolution of star clusters and the chemical evolution of the
Galaxy. A better understanding of the underlying physical processes that may
affect abundance determinations in open clusters is thus essential.

No correlation is found both between the metallicity and mass of the convection
zone, and between the Li and Fe abundances, i.e., no signature of accretion of
H-deficient planetary material is found. Thus our current observations for a
limited sample of IC\,4665 dwarfs seem to favor the scenario that the high
metallicity of SWPs is simply the consequence that planets form more
efficiently in metal-rich environs. However, given that many details of the
processes of star-planet interactions remain not well understood, further
studies are favorable before the problem of the high metallicity of SWPs can be
settled.

Finally, using the deduced standard deviation of [Fe/H] as an upper limit on
the metallicity dispersion amongst the sample stars, we cast various
constraints to show that 1) The total reservoir of heavy elements retained by
their nascent disks is less than five times that of the minimum mass nebula; 2)
The retention efficiency of planet building material is high; 3) The
accumulation of grains may be locally enhanced; 4) Efficient gas accretion may
have initiated around cores with only a few $M_\oplus$; 5) The amount of
protracted accretion of planetary material is limited; and 6) The migration of
gas giants and the circularization of terrestrial planets' orbits are not
regulated by their interaction with a residual population of planetesimals and
dust particles.

\begin{acknowledgements} The authors wish to thank Dr. Debra Fischer for her
expert assistance in the usage of SME. ZXS and XWL acknowledge Chinese NSFC
Grant 10373015. This work is supported by NASA (NAGS5-11779, NNG04G-191G), JPL
(1228184), and NSF (AST-9987417).

\end{acknowledgements}

\end{document}